%% file: robens.tex
\documentclass{moriond}
\usepackage{amssymb,amsmath}     
\usepackage{xspace}

\def\al{\alpha}

\def\be{\begin{equation}}
\def\ee{\end{equation}}
\def\bea{\begin{eqnarray}}
\def\eea{\end{eqnarray}}

\newcommand{\GeV}{{\ensuremath\rm GeV}}
\newcommand{\TeV}{{\ensuremath\rm TeV}}
\newcommand{\pb}{{\ensuremath\rm pb}}
\newcommand{\fb}{{\ensuremath\rm fb}}
\newcommand{\lb}{\left (}
\newcommand{\rb}{\right)}
\newcommand{\lam}{\lambda}
\newcommand{\eqn}{equation}
\newcommand{\Ztwo}{\ensuremath{{\mathbb{Z}_2}}\xspace}
\newcommand{\eqcomma}{\,,}
\newcommand{\eqdot}{\,.}

\newcommand{\Photo}{}

\begin{document}
\rightline{RBI-ThPhys-2022-16}
\vspace*{4cm}
\title{MORE DOUBLETS AND SINGLETS}

\author{T. ROBENS}

\address{Ruder Boskovic Institute, Bijenicka cesta 54, 10000 Zagreb, Croatia}

\maketitle\abstracts{
I give an overview on models that extend the Standard Model scalar sector by additional gauge singlets or multiplets. I discuss current constraints on such models, as well as possible signatures and discovery prospects at current and future colliders.
}

\section{Introduction}
After the discovery of a particle that complies with the properties of the Standard Model (SM) Higgs boson, particle physics has entered an exciting era. One important question is to investigate whether the scalar sector realized in nature corresponds indeed to the SM, or whether it is enhanced by additional scalars, which can be singlets, doublets, or any other multiplets under the electroweak gauge group. Such extensions then typically come with additional particle content, i.e. additional neutral or charged scalars which can also differ by their CP properties. Naturally, such models then need to obey current constraints from both theory and experiment, as e.g. stabilization conditions for the vacuum, positivity, perturbativity, and constraints from direct searches, signal strength, or electroweak precision observables. In turn, in the allowed regions novel signatures can appear that could be of interest for current and future collider searches.

In the following, I discuss several such extensions. For various of these, I present work done by myself and collaborators; for these, we made use of private codes as well as publicly available tools such as HiggsBounds \cite{Bechtle:2020pkv},  HiggsSignals \cite{Bechtle:2020uwn}, 2HDMC \cite{Eriksson:2009ws}, micrOMEGAs \cite{Belanger:2018ccd,Belanger:2020gnr}, and  \texttt{ScannerS}~\cite{Coimbra:2013qq,Ferreira:2014dya,Costa:2015llh,Muhlleitner:2016mzt}. Predictions for production cross sections shown here have been obtained using  Madgraph5 \cite{Alwall:2011uj}.

\section{Real singlet extension}
We first turn to a simple example, where the SM scalar sector has been enhanced by a real singlet field obeying a $\mathbb{Z}_2$ symmetry \cite{Pruna:2013bma,Robens:2015gla,Robens:2016xkb,deFlorian:2016spz,Ilnicka:2018def,DiMicco:2019ngk}. The $\mathbb{Z}_2$ symmetry is softly broken by a vacuum expectation value (vev) of the singlet field, inducing mixing between the gauge-eigenstates which introduces a mixing angle $\al$. The model has in total 5 free parameters. Two of these are fixed by the measurement of the $125\,\GeV$ resonance mass and electroweak precision observables. We then have
\begin{\eqn}
\sin\al,\,m_2,\,\tan\beta\,\equiv\,\frac{v}{v_s}
\end{\eqn}
as free parameters of the model, where $v\,(v_s)$ are the doublet and singlet vevs, respectively. We concentrate on the case where $m_2\,\geq\,125\,\GeV$, where SM decoupling corresponds to $\sin\al\,\rightarrow\,0$.

\begin{center}
\begin{figure}[tbh!]
\begin{center}
\includegraphics[width=0.45\textwidth]{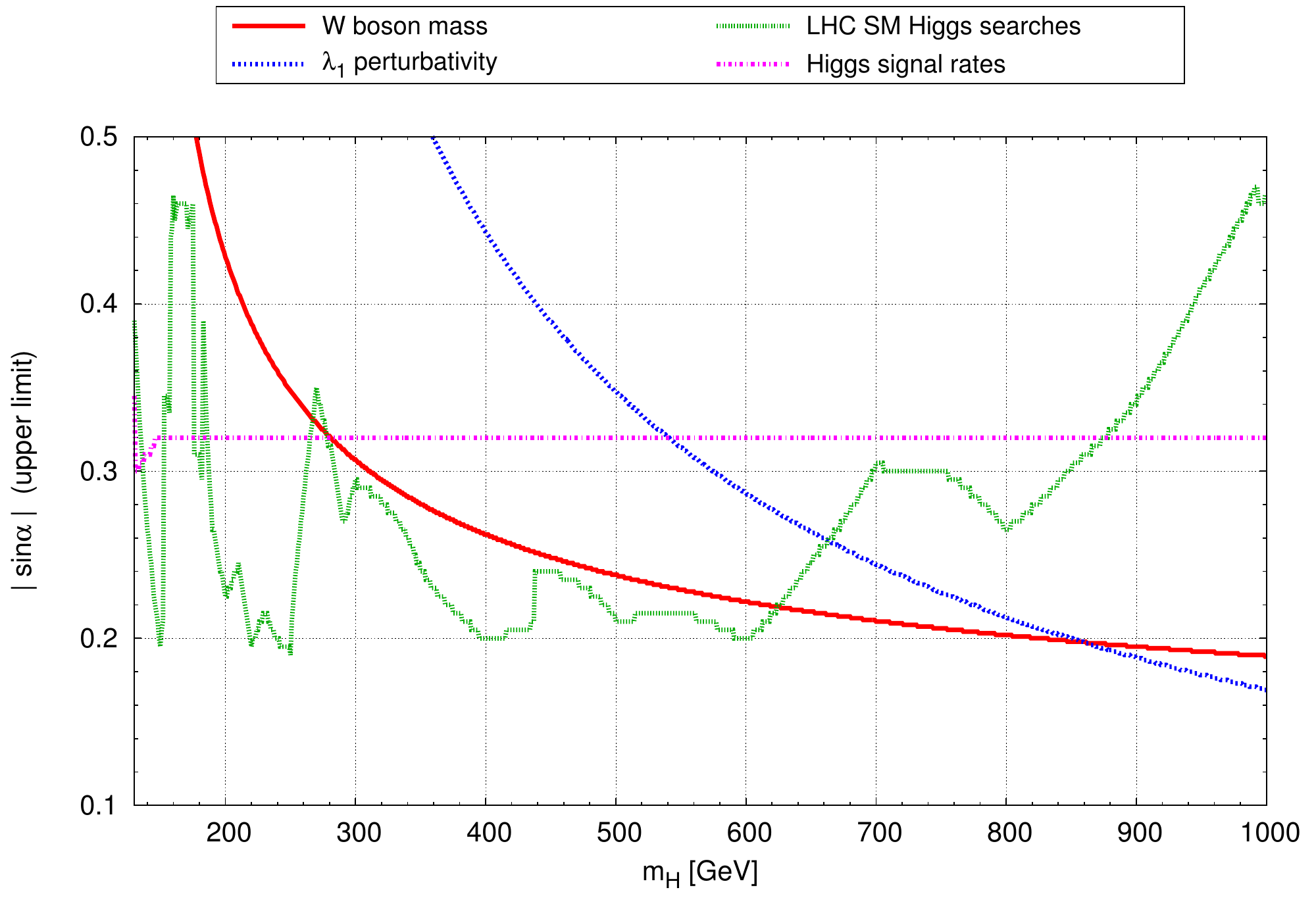}
\includegraphics[width=0.4\textwidth]{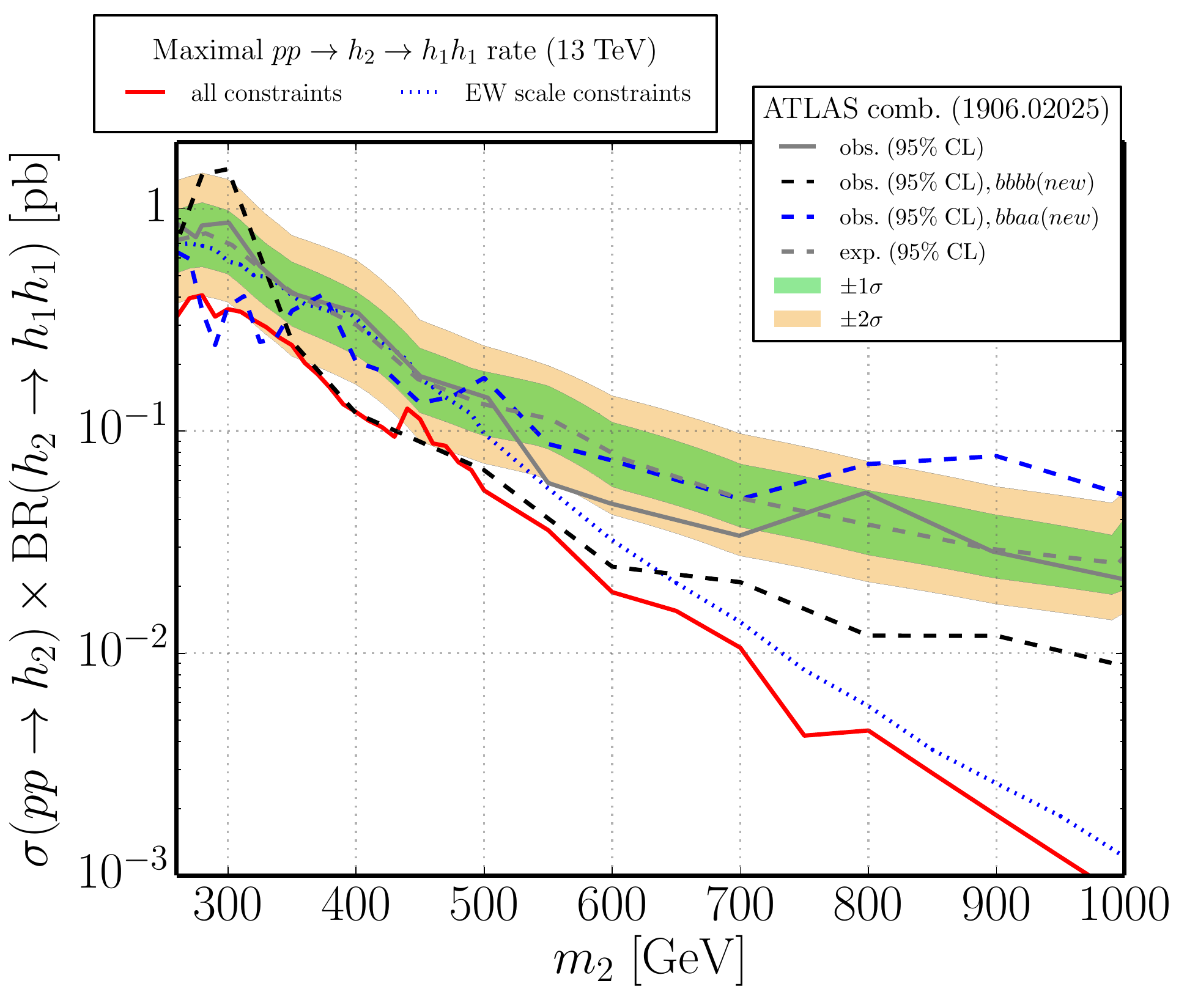}
\caption{\label{fig:singlet} Current constraints on the Higgs singlet extension with a $\mathbb{Z}_2$ symmetry. Constraints have been updated to reflect results prior to Moriond 2022. {\sl Left:} Various constraints on the mixing angle as a function of the second scalar mass, for fixed $\tan\beta=0.1$. {\sl Right:} Current constraints for $H\,\rightarrow\,h_{125} h_{125}$ searches. Comparison with ATLAS Run I combination as well as novel results (see text for details).}
\end{center}
\end{figure}
\end{center}

The model is subject to a number of theoretical and experimental constraints. Exemplary limits are shown in figure \ref{fig:singlet}, which correspond to an update to results presented in \cite{Robens:2021rkl}, including a comparison of the currently maximal available rate of $H\,\rightarrow\,h_{125}h_{125}$ with the combination limits from ATLAS \cite{Aad:2019uzh} as well as novel results for $b\bar{b}b\bar{b}$ \cite{ATLAS:2022hwc} and $b\bar{b}\gamma\gamma$ \cite{ATLAS:2021ifb} final states. The most constraining direct search bounds are in general dominated by searches for diboson final states \cite{CMS-PAS-HIG-13-003,Khachatryan:2015cwa,Sirunyan:2018qlb,Aaboud:2018bun}. In some regions, the Run 1 Higgs combination \cite{CMS-PAS-HIG-12-045} is also important. Especially \cite{Sirunyan:2018qlb,Aaboud:2018bun} currently correspond to the best probes of the models parameter space\footnote{We include searches currently available via HiggsBounds.}.
\section{Two Higgs Doublet Models}

Two Higgs doublet models (2HDMs) constitute another example of new physics models introducing additional scalar states. A general discussion of such models is e.g. given in \cite{Branco:2011iw} and will not be repeated here. In these models, the SM scalar sector is augmented by a second scalar doublet which also acquires a vev; electroweak symmetry breaking then involves both fields. Several structures of couplings in the Yukawa sector are possible and distinguish the different models. The particle content in the scalar sector containa, besides the SM candidate, two additional neutral scalars which differ in CP properties as well as a charged scalar, denoted by $h,\,H,\,A,\,H^\pm$, where one of the two CP-even neutral scalars $h,\,H$ needs to be identified with the 125 \GeV~ resonance discovered at the LHC. Besides the masses of the scalar particles, the scalar sector is also characterized by different mixing angles, which are typically parametrized in terms of $\cos\lb \beta-\alpha\rb$ and $\tan\beta$.

\begin{center}
\begin{figure}[tbh!]
\begin{center}
\includegraphics[width=0.4\textwidth]{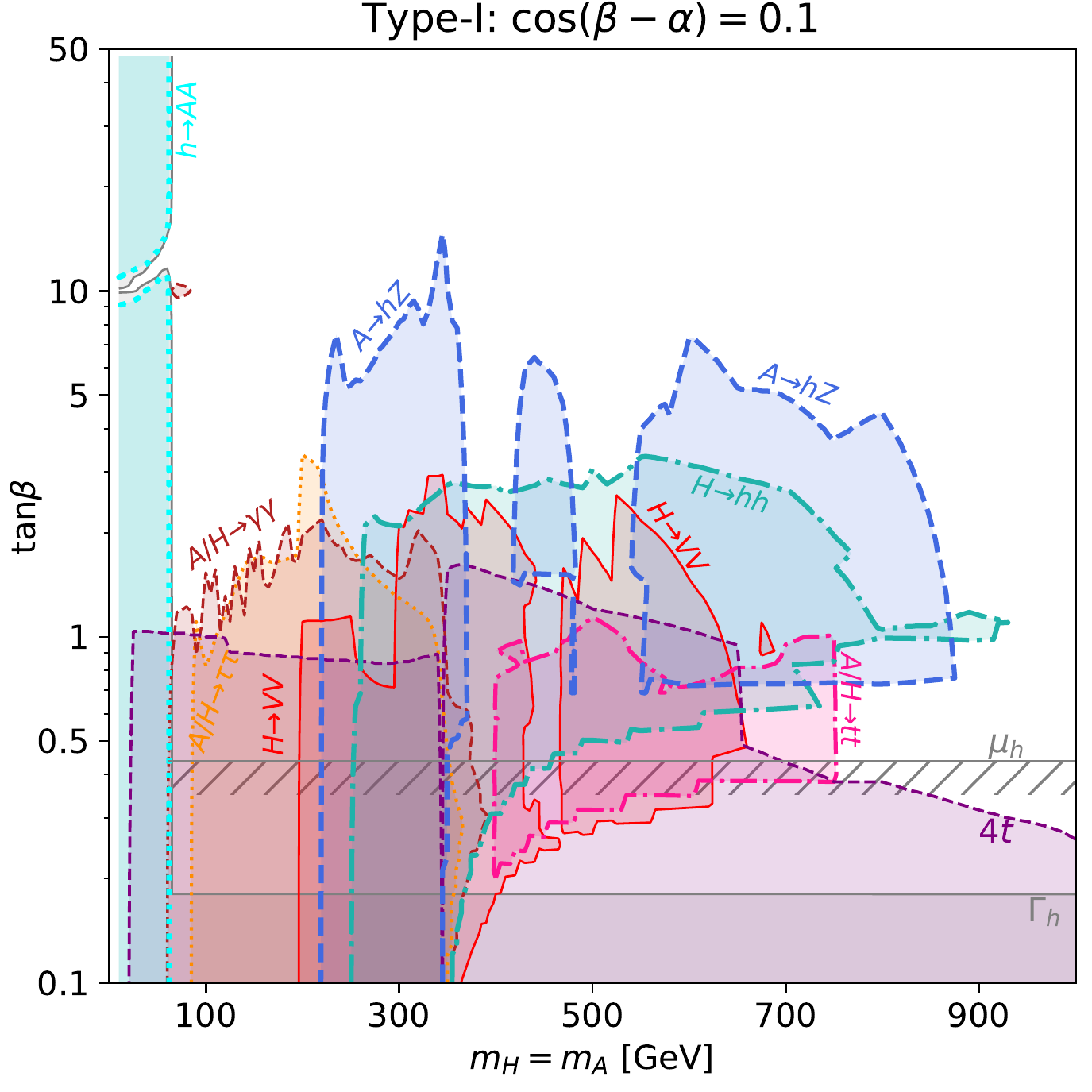}
\includegraphics[width=0.4\textwidth]{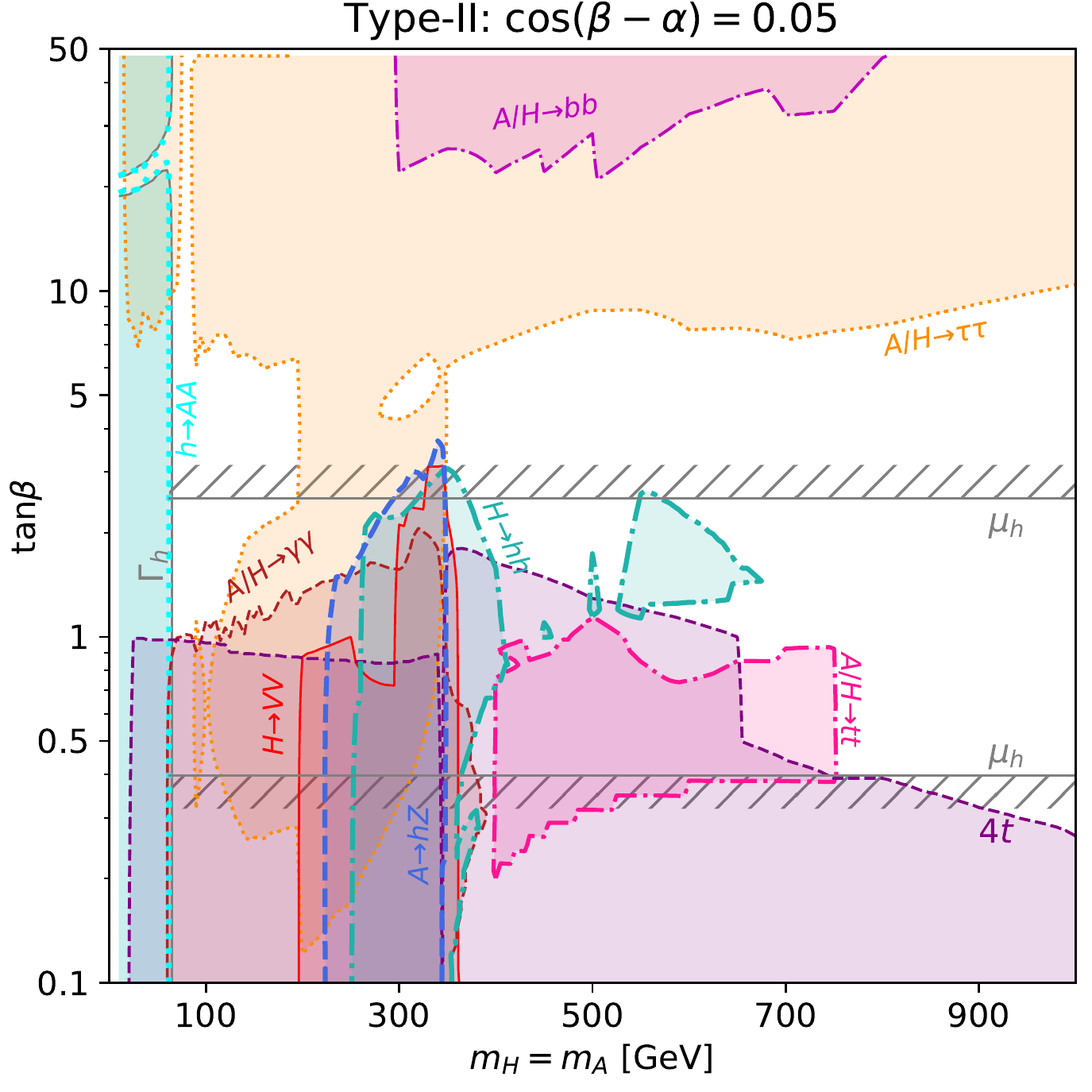}
\caption{\label{fig:2hdmsearches} Constraints on various types of 2HDM models, from various search channels at the LHC, status 2020. See text for reference.}
\end{center}
\end{figure}
\end{center}

Exemplarily, in figure \ref{fig:2hdmsearches} we show constraints on two different types of 2HDMs from various searches, taken from \cite{Kling:2020hmi}, for the degenerate case where all heavy scalar masses are set equal. In that work, the authors considered various searches at the 8 \TeV~ and early 13 \TeV~ LHC runs. We refer to the original work for a complete list of all channels.

\begin{center}
\begin{figure}[tbh!]
\begin{center}
\includegraphics[width=0.4\textwidth]{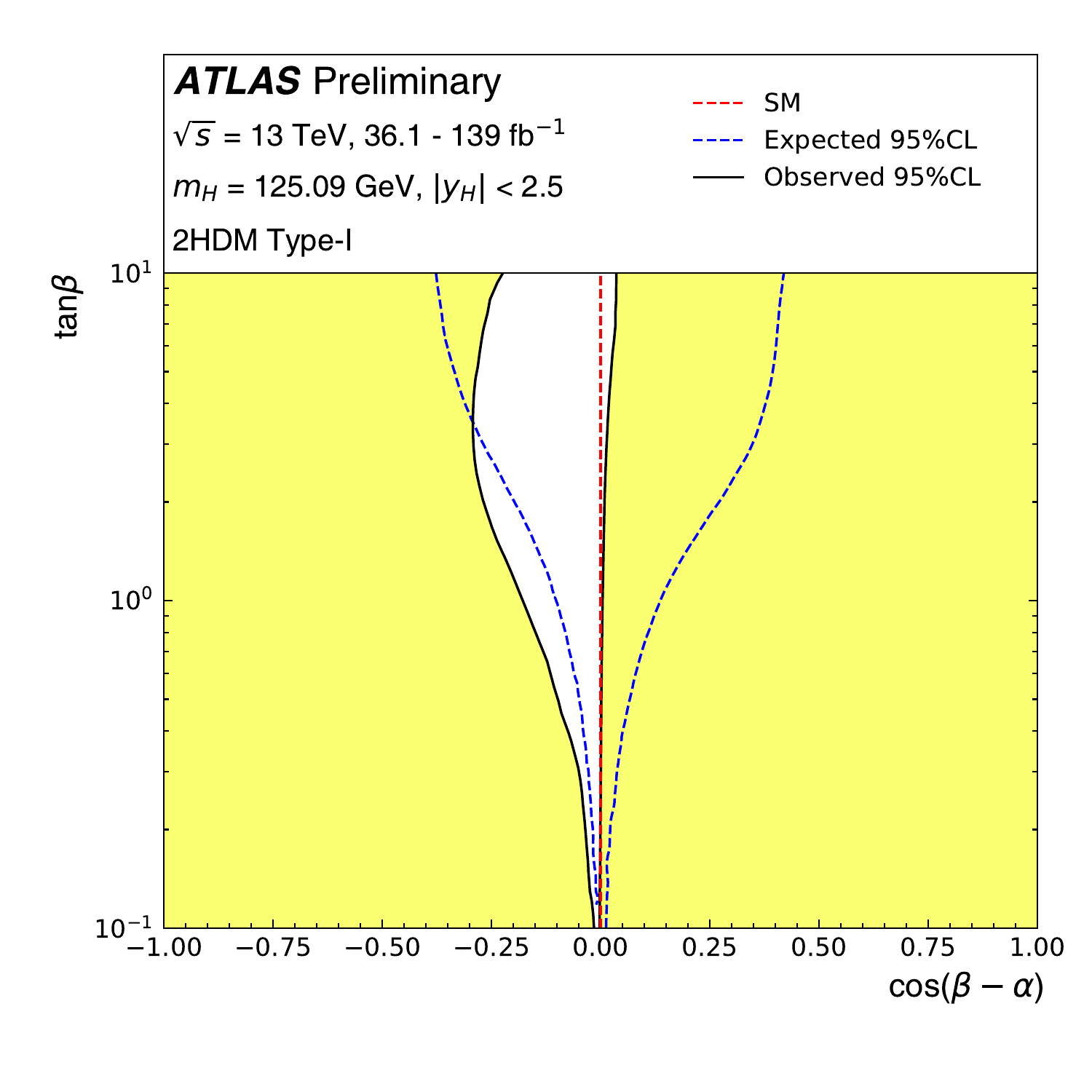}
\includegraphics[width=0.4\textwidth]{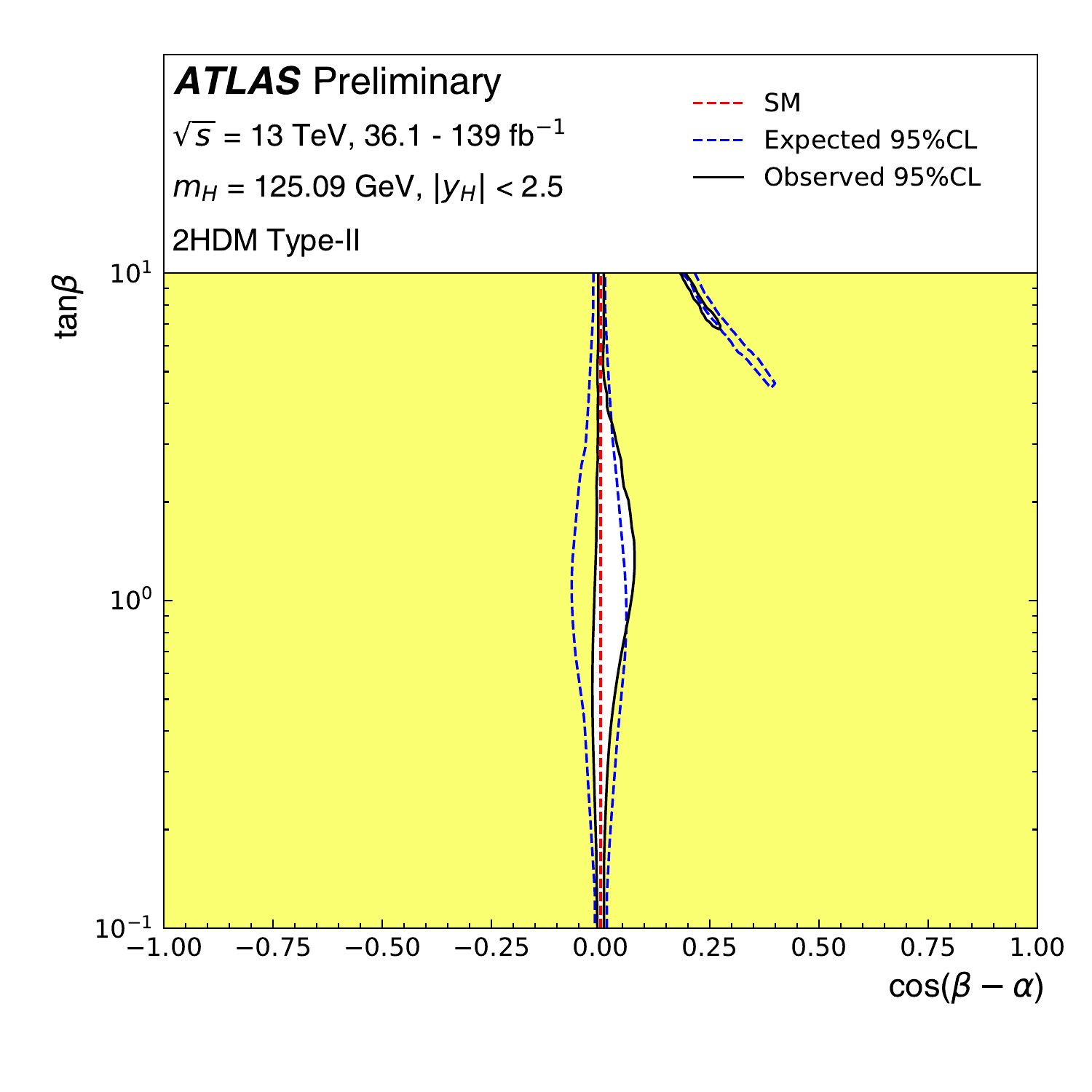}
\includegraphics[width=0.4\textwidth]{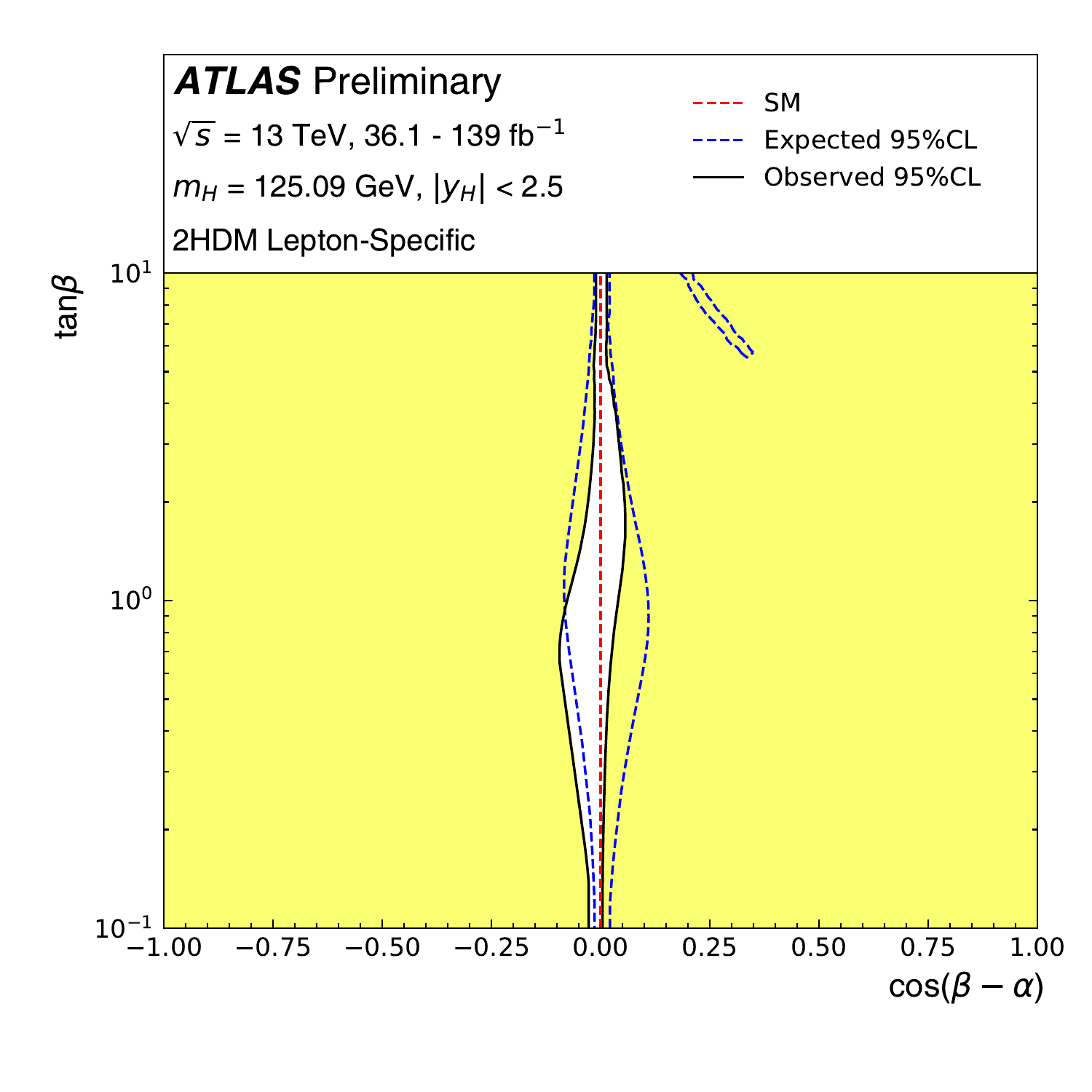}
\includegraphics[width=0.4\textwidth]{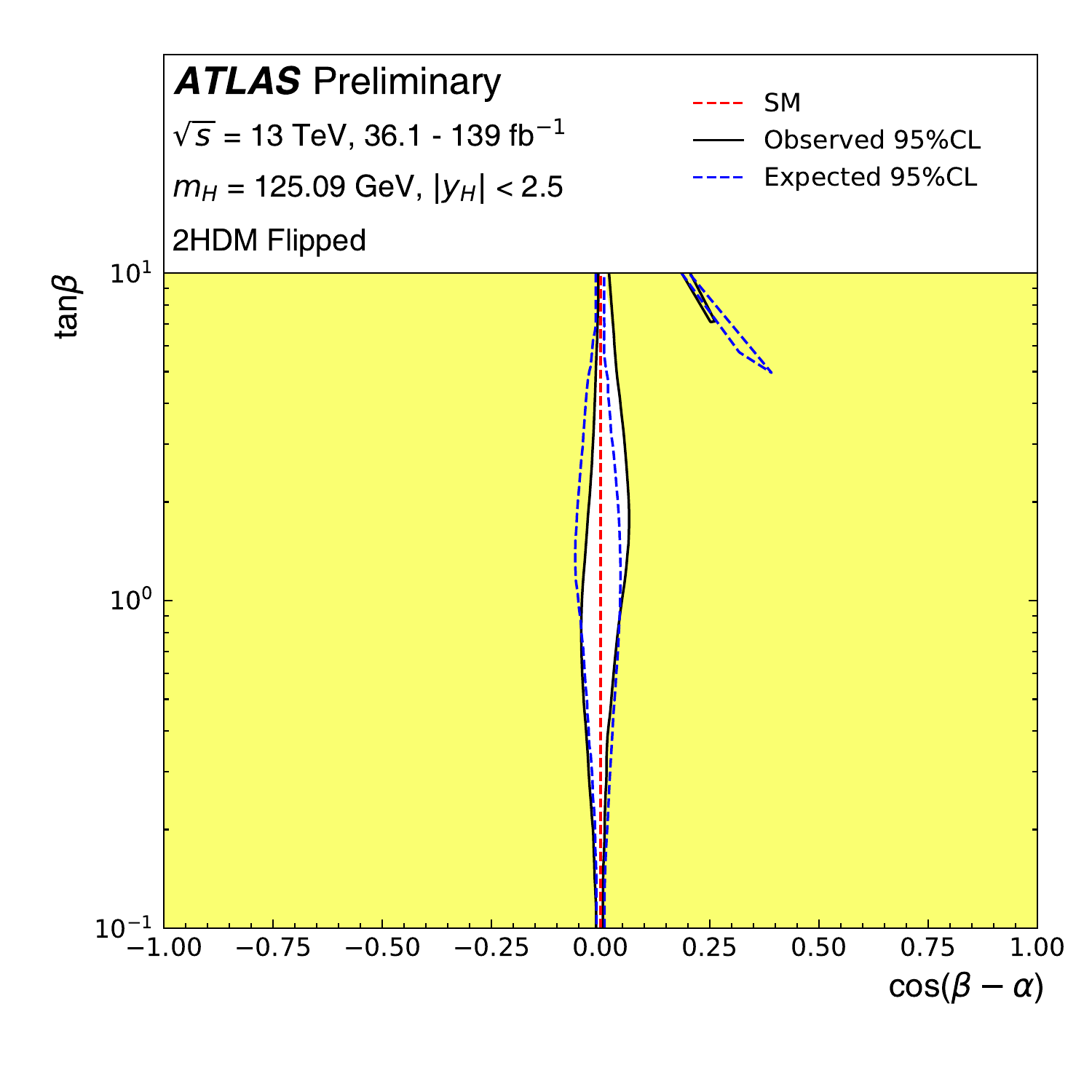}
\caption{\label{fig:atlcba} Signal strength fit constraints and allowed regions for $\cos\lb\beta-\alpha\rb$ for various types of 2HDMs; preliminary ATLAS combination results. See text for reference.}
\end{center}
\end{figure}
\end{center}

Another important result is that in 2HDMs, the parameter space for the mixing angle $\cos\lb\beta-\alpha\rb$ is stronlgy constrained from signal strength measurements, where the exact allowed region again depends on the Yukawa structure of the model. In figure \ref{fig:atlcba}, we display the results as obtained by the ATLAS collaboration in a fit combination using full Run 2 data \cite{ATLAS-CONF-2021-053}. The actual constraints differ for each model, but the absolute value of the mixing angle does not exceed $0.25$.
\section{Two real scalar extension at hadron colliders}

\input{2rscalars}

\begin{center}
\begin{figure}[tbh!]
\begin{center}
\includegraphics[width=0.45\textwidth]{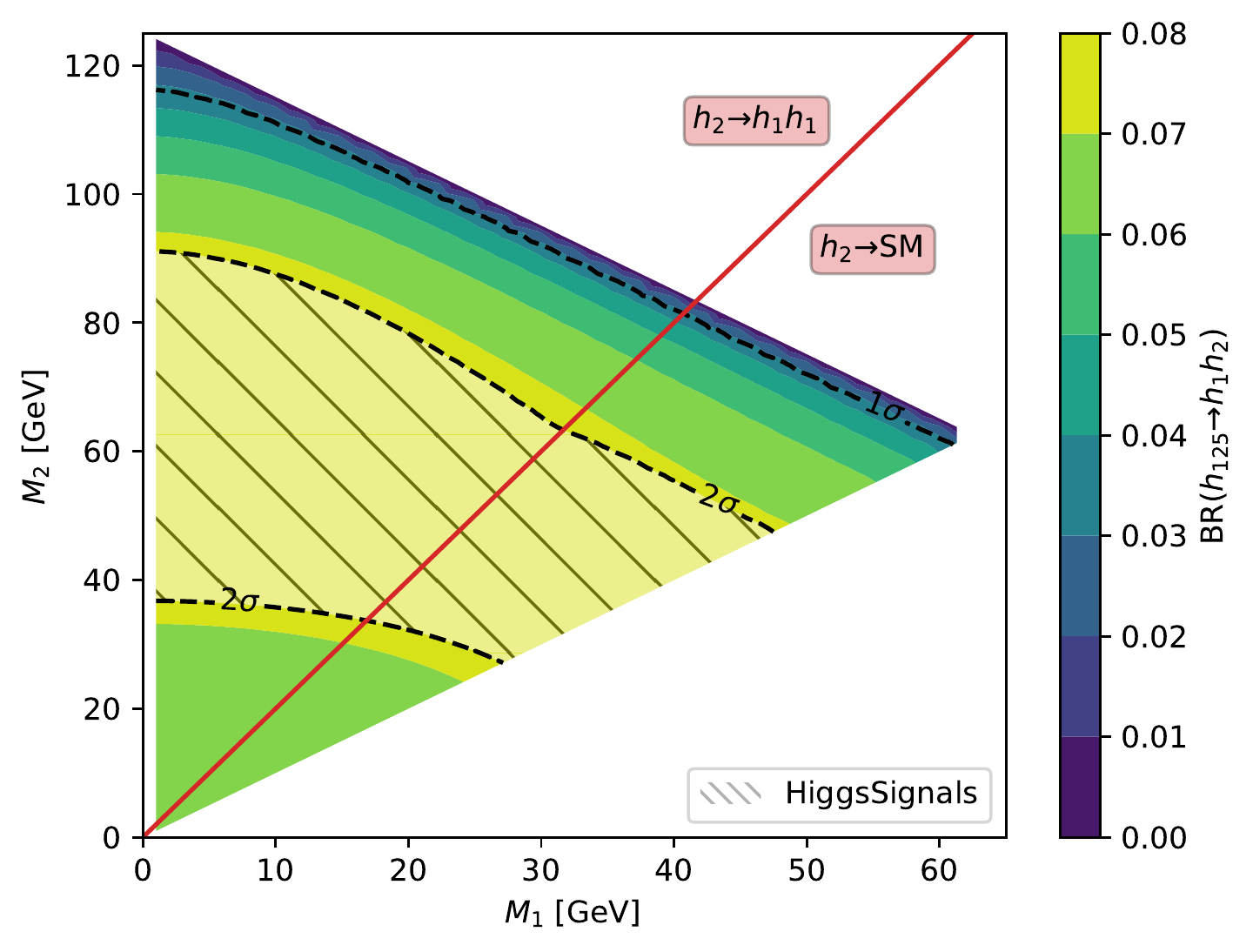}
\includegraphics[width=0.45\textwidth]{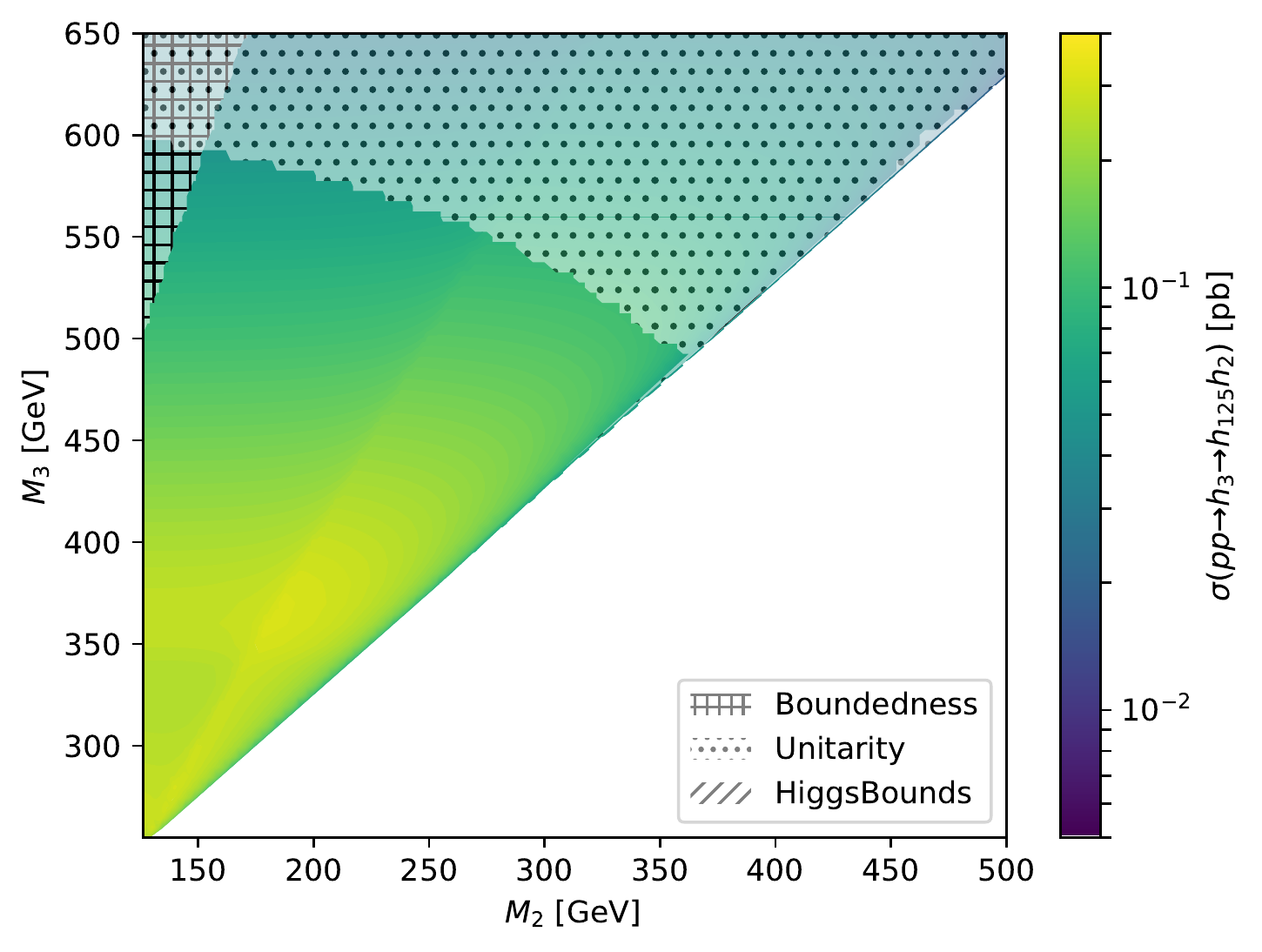}
\includegraphics[width=0.45\textwidth]{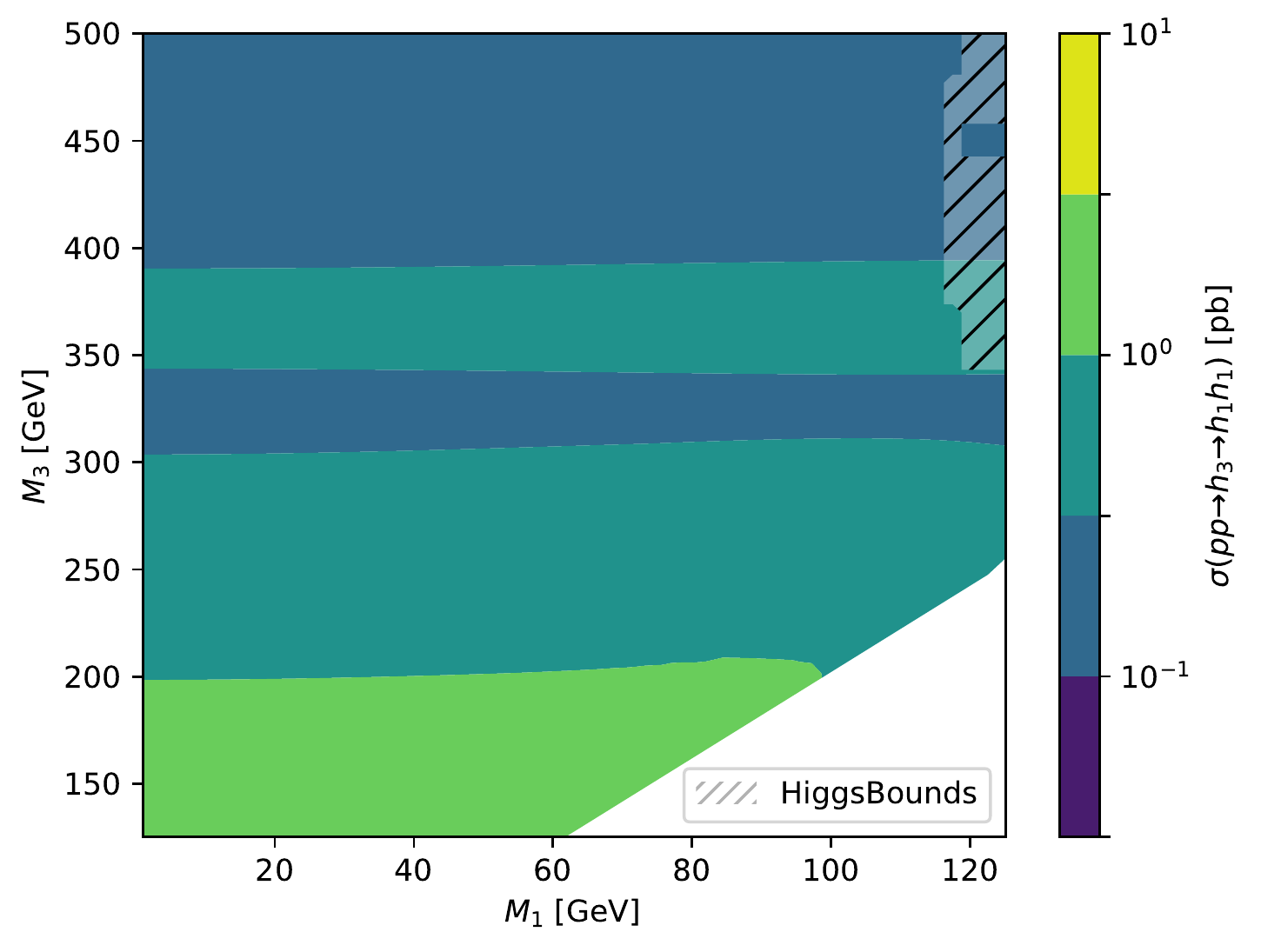}
\includegraphics[width=0.45\textwidth]{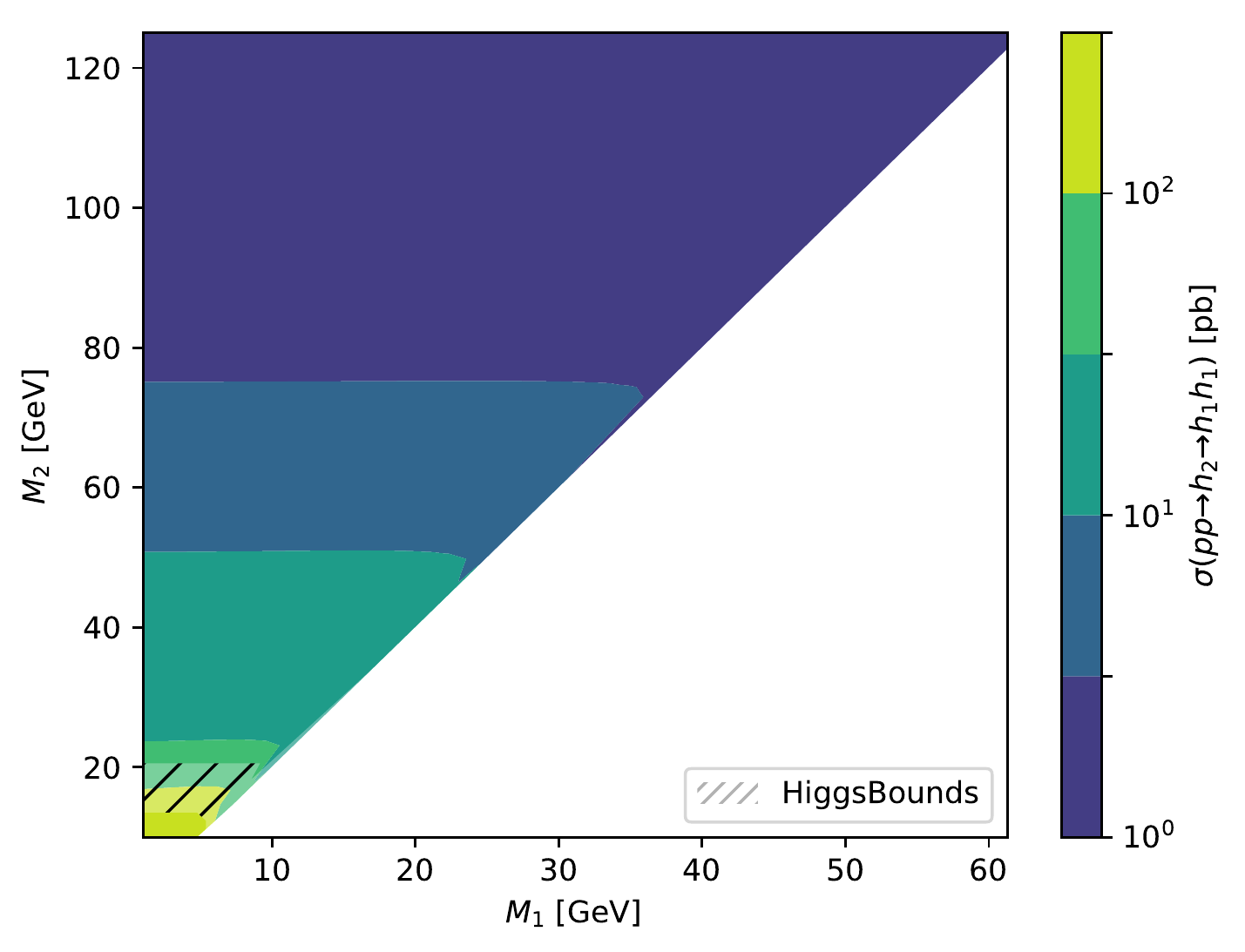}
\caption{\label{fig:2rbps} Benchmark planes for the TRSM, for asymmetric (top) and symmetric (bottom) final states, as defined in the text. {\sl Top left:} BP1, where $h_3\,\equiv\,h_{125}$; displayed is the branching ratio $h_{125}\,\rightarrow\,h_1\,h_2$ in the two-dimensional mass plane. {\sl Top right:} BP3; {\sl Bottom left:} BP5; {\sl Bottom right:} BP6. For the last three, production cross sections at a 13 \TeV~ $pp$ collider are shown. Exclusion bounds on these planes from various constraints are also given.}

\end{center}
\end{figure}
\end{center}
\section{Inert Doublet Model}
\begin{center}
\begin{figure}[tbh!]
\begin{center}
\includegraphics[width=0.45\textwidth]{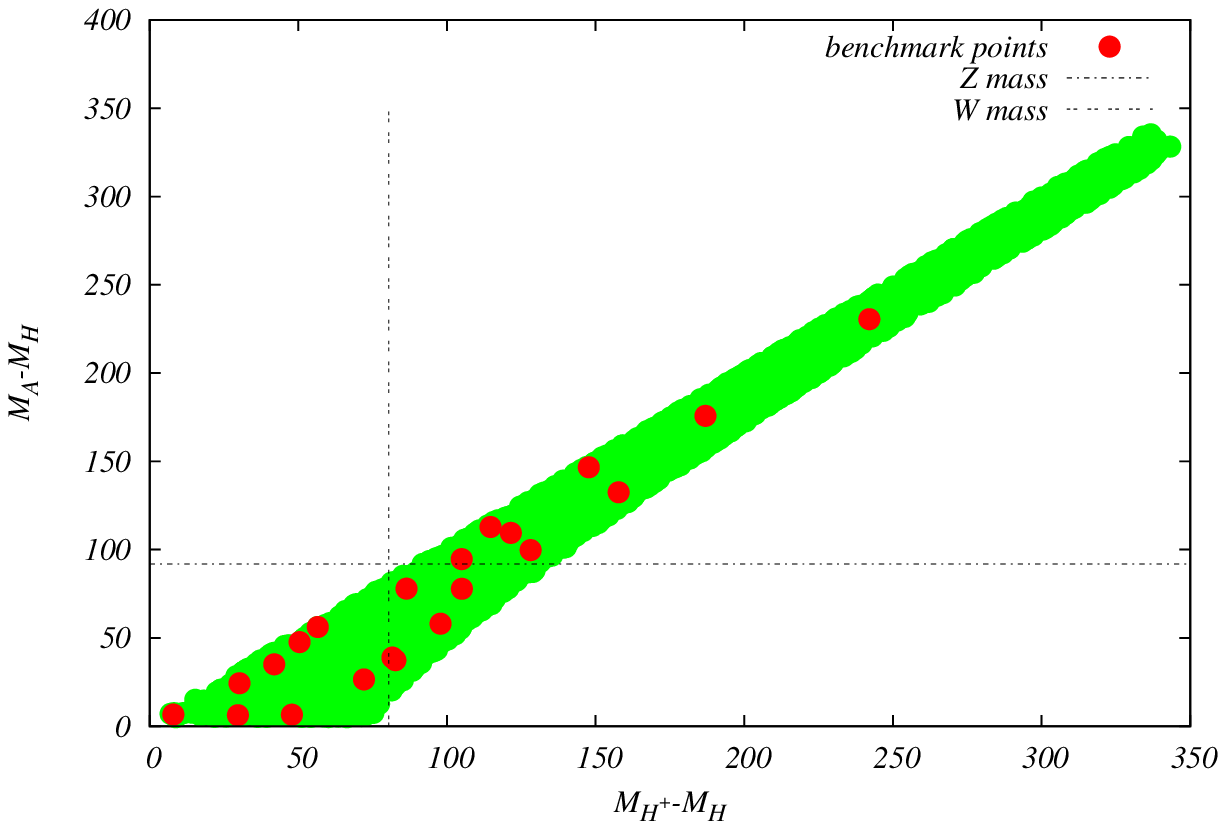}
\includegraphics[width=0.45\textwidth]{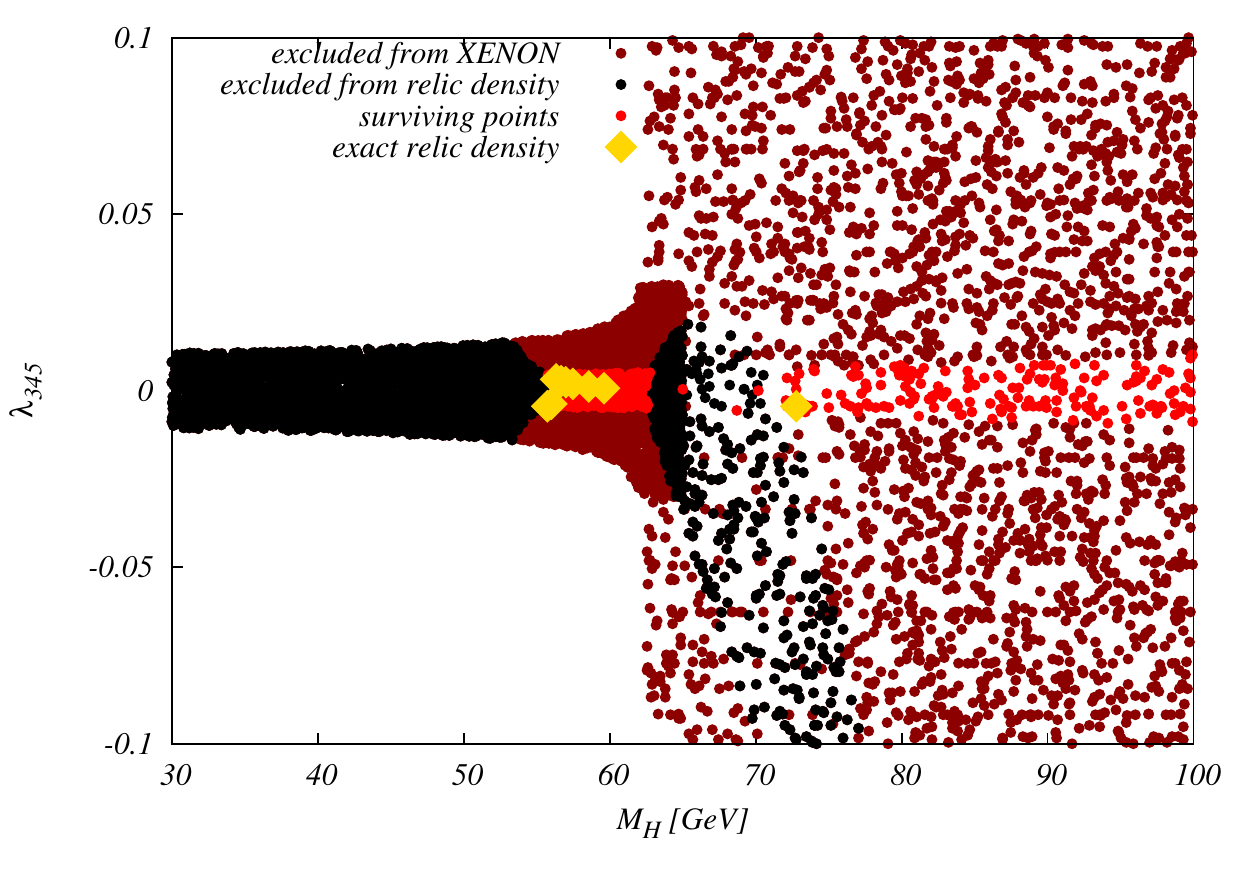}
\caption{\label{fig:idmscan} {\sl Left:} allowed range in the $(M_{H^\pm}-M_H, M_A-M_H)$ plane, including benchmark points discussed in the text. The lines discriminate on- off-shellness of the decay products of $H^\pm$ and $A$ respectively. {\sl Right:} allowed and excluded regions in the $M_H,\lam_{345}$ parameter space. Red points are still allowed after all constraints, golden render exact relic density.}
\end{center}
\end{figure}
\end{center}

We now turn to a new physics model that contains a dark matter candidate.
The Inert Doublet Model is a two Higgs doublet model with an exact discrete $\mathbb{Z}_2$ symmetry \cite{Deshpande:1977rw,Cao:2007rm,Barbieri:2006dq}. The model contains four additional scalar states $H,\,A,\,H^\pm$, and has in total 7 free parameters prior to electroweak symmetry breaking:
$v,\,m_h,\,\underbrace{m_H,\,m_A,\,m_{H^\pm}}_{\text{second doublet}},\,\lam_2,\,\lam_{345}\,\equiv\,\lam_3+\lam_4+\lam_5$.
Here, the $\lam_i$s denote standard couplings appearing in the 2HDM potential. Two parameters ($m_h$ and $v$) are fixed by current measurements. We have provided a detailed investigation of the models parameter space in \cite{Ilnicka:2015jba,Ilnicka:2018def,Kalinowski:2018ylg,Kalinowski:2020rmb,Robens:2021yrl}. One important observation is the existance of a relatively strong degeneracy between the additional masses of the second doublet, as well as a minimal mass scale for the dark matter candidate $H$ resulting from a combination of relic density and signal strength measurement constraints (see \cite{Ilnicka:2015jba,Kalinowski:2020rmb} for a detailed discussion). These features are displayed in figure \ref{fig:idmscan}.

\begin{center}
\begin{figure}[tbh!]
\begin{center}
\includegraphics[width=0.45\textwidth]{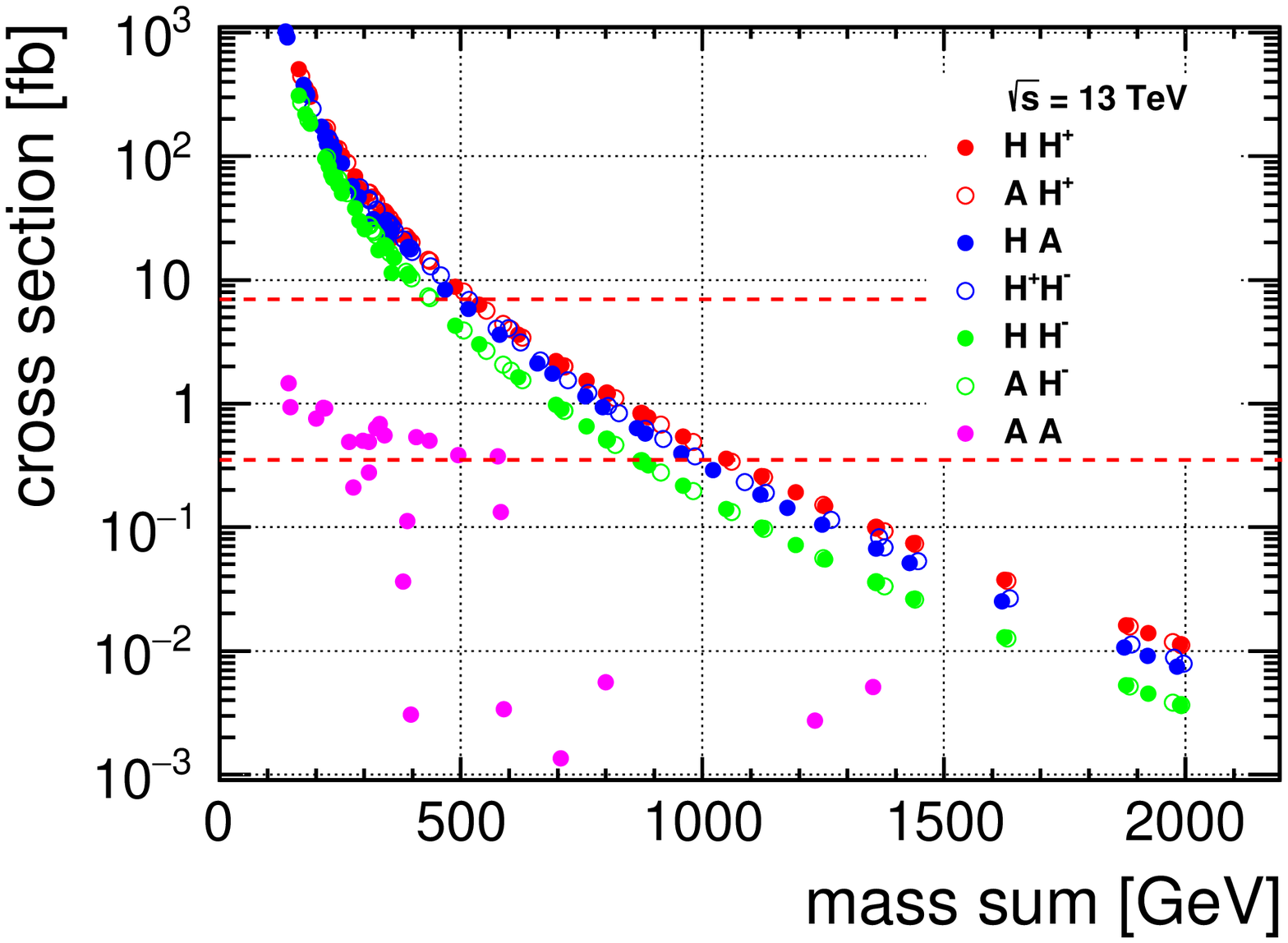}
\includegraphics[width=0.45\textwidth]{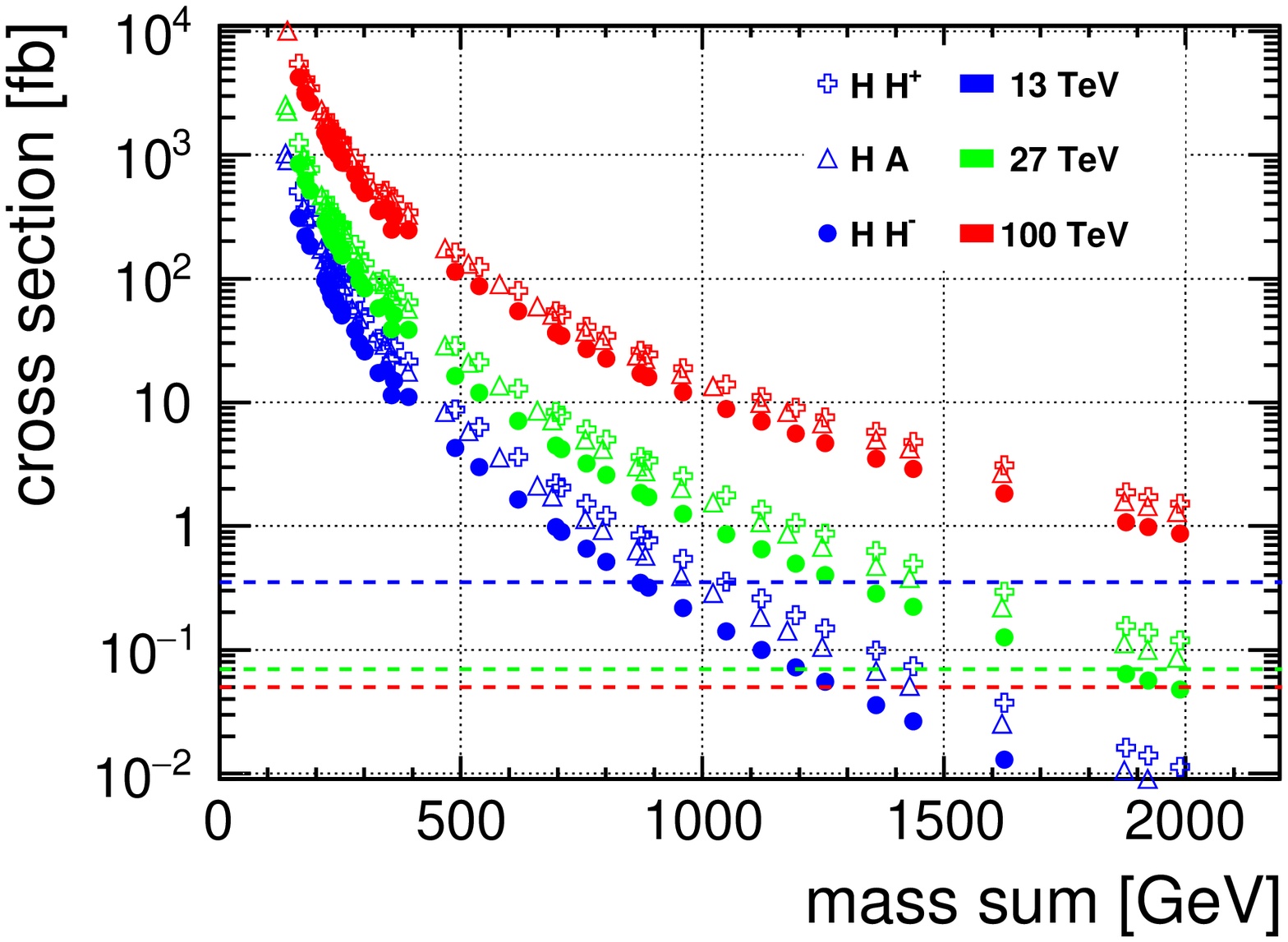}
\includegraphics[width=0.45\textwidth]{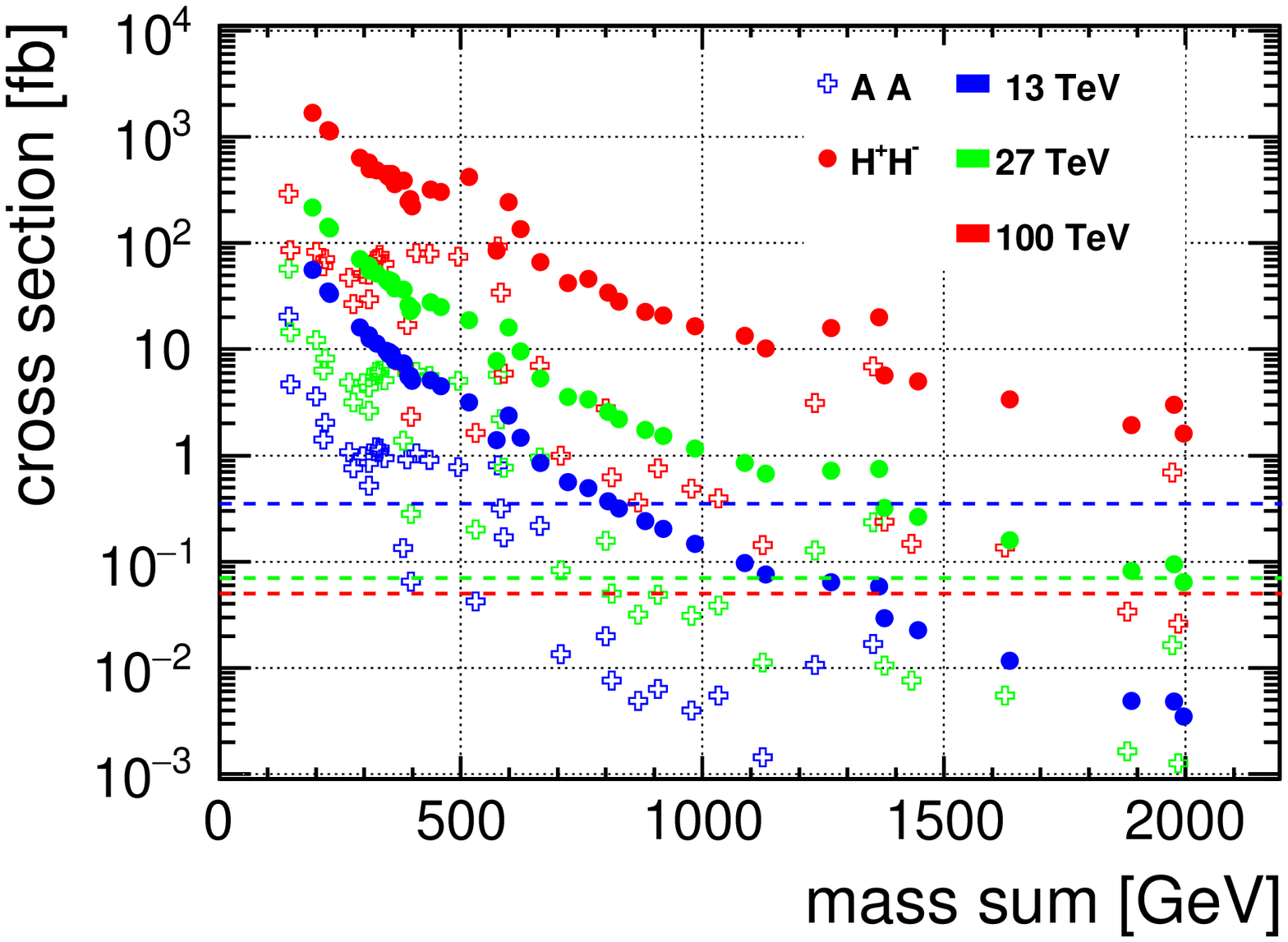}
\includegraphics[width=0.45\textwidth]{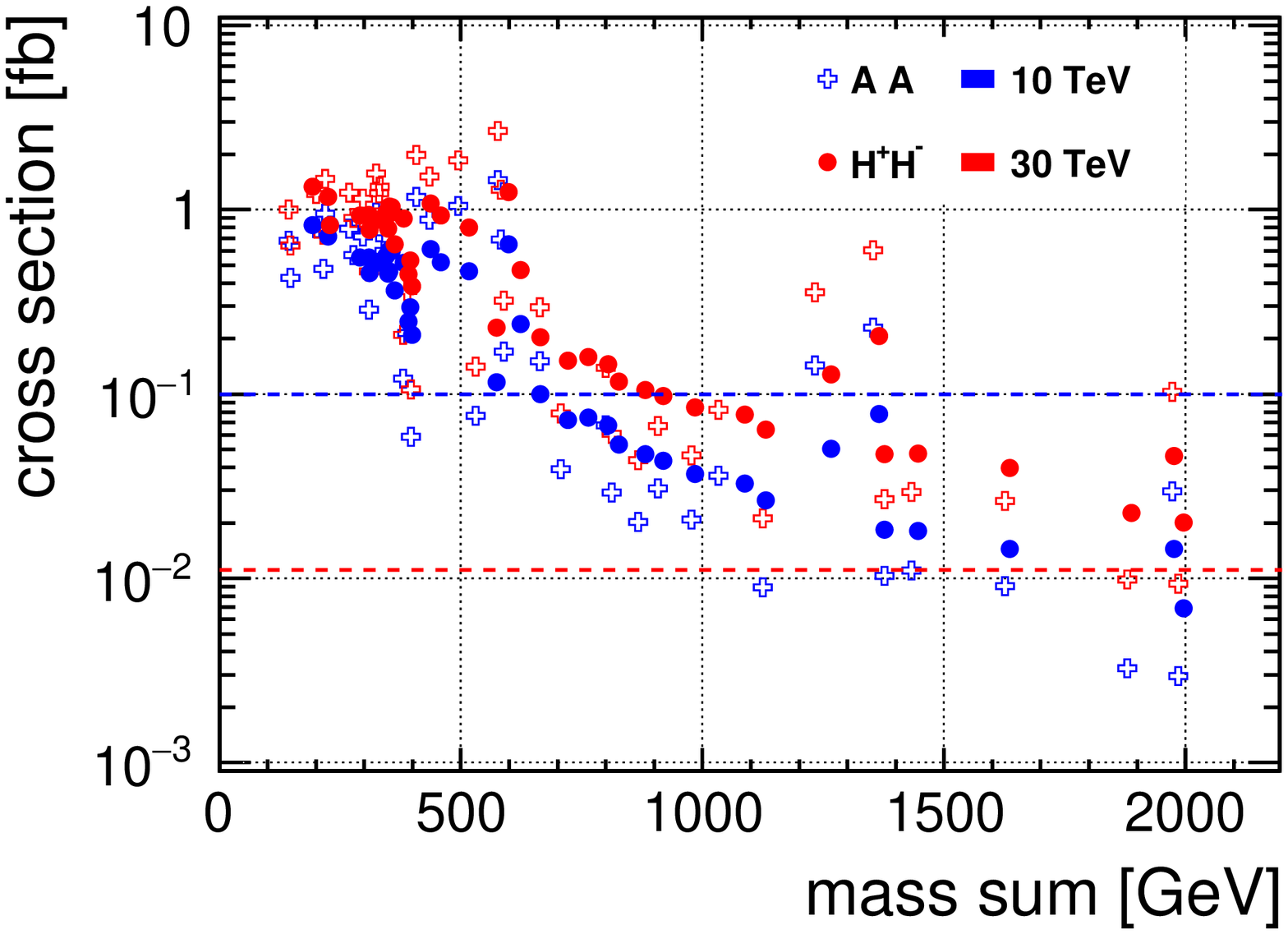}
\caption{\label{fig:idm} Predictions for production cross sections for various processes and collider options. {\sl Top left:} Predictions for various pair-production cross sections for a $pp$ collider at 13 \TeV, as a function of the mass sum of the produced particles. {\sl Top right:} Same for various center-of-mass energies. {\sl Bottom left:} VBF-type production of $AA$ and $H^+\,H^-$ at various center-of-mass energies for $pp$ colliders. {\sl Bottom right:} Same for $\mu^+\mu^-$ colliders. The lines correspond to the cross-sections required to produce at least 1000 events using the respective design luminosity.}
\end{center}
\end{figure}
\end{center}

In \cite{Kalinowski:2020rmb}, we have performed a sensitivity comparison for selected benchmark points \cite{Kalinowski:2018ylg,Kalinowski:2018kdn,Kalinowski:2020rmb}, relying on a simple counting criteria: a benchmark point is considered reachable if at least 1000 signal events are produced using nominal luminosity of the respective collider. The corresponding results in terms of mass scales of the pair-produced particles are displayed in table \ref{tab:sens}, with accompagnying plots in figure \ref{fig:idm}, taken from \cite{Kalinowski:2020rmb}. We here have used Madgraph5 \cite{Alwall:2011uj} with a UFO input file from \cite{Goudelis:2013uca} for cross-section predictions. In contrast, results for CLIC have been obtained using a detailed study of signal and background \cite{Kalinowski:2018kdn,CLIC:2018fvx}.
\begin{center}
\begin{table}[tbh!]
\begin{center}
\begin{tabular}{||c||c||c||c||} \hline \hline
{collider}&{ all others}& {AA} & {AA +VBF}\\ \hline \hline
HE-LHC&2 \TeV&400-1400 \GeV&800-1400 \GeV\\
FCC-hh&2 \TeV&600-2000 \GeV&1600-2000 \GeV\\ \hline \hline
CLIC, 3 \TeV&2 \TeV &- &300-600 \GeV\\
$\mu\mu$, 10 \TeV&2 \TeV &-&400-1400 \GeV\\
$\mu\mu$, 30 \TeV&2 \TeV &-&1800-2000 \GeV \\ \hline \hline
\end{tabular}
\caption{\label{tab:sens} Sensitivity of different collider options, using the sensitivity criterium of 1000 generated events in the specific channel. $x-y$ denotes minimal/ maximal mass scales that are reachable.}
\end{center}
\end{table}
\end{center}
\section{Conclusions}
I have presented results for various models that extend the scalar sector of the SM by additional gauge singlets or doublets. These models are subject to additional theoretical and experimental constraints limiting the parameter space. On the other hand, all of these allow for additional signatures that have not fully been explored yet at current colliders, and therefore provide encouragement for the experimental collaborations to investigate these at future hadron or lepton machines. Despite the discovery of a particle that complies with the properties of a SM Higgs, the structure of the electroweak potential is not completely understood yet. The models discussed here will allow to further investigate the electroweak structure realized in nature.
\section*{Acknowledgements}
The author wants to thank the organizers for the invitation, a very pleasant atmosphere, as well as financial support. I furthermore thank all collaborators who helped to achieve the results presented here. 
\section*{References}

\input{robens.bbl}
\end{document}

%% file: 2rscalars.tex
We now turn to the two-real-singlet extension (TRSM) \cite{Robens:2019kga}, a model that extends the SM scalar sector by two real scalars that are singlets under the SM gauge group. This model allows for interesting new final states, as e.g. scalar-to-scalar decays. Single scalar production following by asymmetric scalar decays or symmetric decays, where all scalar masses differ from the SM-like scalar mass at 125 \GeV, have not yet been fully explored. Simple counting reveals that for such scenarios at least three physical scalar states need to be present in the model, out of which one takes the role of the state already discovered by the LHC experiments.

The potential in the scalar sector is given by
\begin{equation}
    \begin{aligned}
        V\lb \Phi,\,S,\,X\rb & = \mu_{\Phi}^2 \Phi^\dagger \Phi + \lambda_{\Phi} {(\Phi^\dagger\Phi)}^2
        + \mu_{S}^2 S^2 + \lambda_S S^4
        + \mu_{X}^2 X^2 + \lambda_X X^2                                              \\
          & \quad+ \lambda_{\Phi S} \Phi^\dagger \Phi S^2
        + \lambda_{\Phi X} \Phi^\dagger \Phi X^2
        + \lambda_{SX} S^2 X^2\eqdot
    \end{aligned}\label{eq:TRSMpot}
\end{equation}
where $\Phi$ denotes the doublet also present in the SM potential and $X,\,S$ are the two additional real scalars. The model obeys an additional $\Ztwo\,\otimes\,\Ztwo'$ symmetry
$        \Ztwo^S: \, S\to -S\eqcomma
        \Ztwo^X: \, X\to -X,$ while all other fields transform evenly under the respective $\Ztwo$ symmetry. All three scalars acquire a vev and mix. This leads to three physical states with all possible scalar-scalar interactions.

In the following, we will use the convention that
\begin{\eqn}\label{eq:hier}
M_1\,\leq\,M_2\,\leq\,M_3
\end{\eqn}
and denote the corresponding physical mass eigenstates by $h_i$.
Gauge and mass eigenstates are related via a mixing matrix. Interactions with SM particles are then inherited from the scalar excitation of the doublet via rescaling factors $\kappa_i$, such that $g_i^{h_i A B}\,=\,\kappa_i\,g_i^{h_i A B,\text{SM}}$ for any $h_i A B$ coupling, where $A,\,B$ denote SM particles. Orthogonality of the mixing matrix implies $\sum_i \kappa_i^2\,=\,1$. 

The model allows for interesting scalar-to-scalar decays
\begin{align}
    pp\rightarrow h_a~(+ X)\rightarrow h_b h_b~(+ X),  \label{eq:process_sym} \\
    pp\rightarrow h_3~(+ X)\rightarrow h_1 h_2~(+ X), \label{eq:process_asym}
\end{align}
where ${a,b}\,\in\,\left\{1,2,3 \right\}$. In \cite{Robens:2019kga}, six benchmark planes (BPs) were suggested for the decay chains above: three involving asymmetric decays (\ref{eq:process_asym}) and three symmetric decays (\ref{eq:process_sym}), where in the latter case we concentrated on scenarios where $M_{a,b}\,\neq\,125\,\GeV$. Depending on the specific benchmark point, production rates can reach $\mathcal{O}\lb 60\,\pb \rb$ for the above production modes. The benchmark scenario which considers $h_3\,\rightarrow\,h_2\,h_2$, where $M_2\,>\,250\,\GeV$, can furthermore lead to interesting $h_{125}h_{125}h_{125}$ and $h_{125}h_{125}h_{125}h_{125}$ final states. Largest rates for these processes, including further decays to SM final states, are in the $\mathcal{O}\lb \fb \rb$ range.

Figure \ref{fig:2rbps} shows predicted production rates for several of these benchmark planes: BPs 1/3 for the asymmetric decay, where $h_{125}\,\equiv\,h_3/ h_1$, as well as two symmetric scenarios BPs 4/ 5 where a non-SM like scalar decays into light scalars with $M_1\,\leq\,125\,\GeV$. Depending on the benchmark point, dominant final states typically contain multiple b-jets $b\bar{b}b\bar{b}\lb b\bar{b}\rb$. While some of these are already under investigation by the LHC experiments, I encourage the collaborations to explore all possible channels and collider signatures.  These can lead to sizeable rates at the 13 \TeV~ LHC, while still complying with all current theoretical and experimental bounds.